\documentclass[11pt, secnumarabic, amssymb, nobibnotes]{revtex4-1}

\usepackage[english]{babel}

\usepackage{graphicx}
\usepackage{pslatex}
\usepackage{xspace}
\usepackage{subfigure}
\usepackage{hyperref}
\usepackage{amsmath}
\usepackage{setspace}
\usepackage[usenames,dvipsnames]{color}
\usepackage{booktabs}

\begin{document}


\newcommand{\marked}{\textcolor{red}}

\title{Influence of phase segregation on the recombination dynamics \\in organic bulk heterojunction solar cells}


\author{A.~Baumann$^1$}

\author{T.~J.~Savenije$^{1,3}$}

\author{D.~H.~K.~Murthy$^3$}

\author{M. Heeney$^4$}

\author{V.~Dyakonov$^{1,2}$}\email{dyakonov@physik.uni-wuerzburg.de}

\author{C.~Deibel$^1$}\email{deibel@physik.uni-wuerzburg.de}
\affiliation{$^1$ Experimental Physics VI, Julius-Maximilians-University of W\"urzburg, 97074 W\"urzburg, Germany}
\affiliation{$^2$ Bavarian Center for Applied Energy Research e.V. (ZAE Bayern), 97074 W\"urzburg, Germany}
\affiliation{$^3$ Optoelectronic Materials Section, Department of Chemical Engineering, Delft University of Technology, 2628 BL Delft, Netherlands}
\affiliation{$^4$ Department of Chemistry, Imperial College London, London, SW7 2AZ, United Kingdom}
\date{\today}


\begin{abstract}

We studied the recombination dynamics of charge carriers in organic bulk heterojunction solar cells made of the blend system poly(2,5-bis(3-dodecyl thiophen-2-yl) thieno[2,3-b]thiophene) (pBTCT-C$_{12}$):[6,6]-phenyl-C$_{61}$-butyric acid methyl ester (PC$_{61}$BM) with a donor--acceptor ratio of 1:1 and 1:4. The techniques of charge carrier extraction by linearly increasing voltage (photo-CELIV) and, as local probe, time-resolved microwave conductivity (TRMC) were used.
We observed a difference in the initially extracted charge carrier concentration in the photo-CELIV experiment by one order of magnitude, which we assigned to an enhanced geminate recombination due to a fine interpenetrating network with isolated phase regions in the 1:1 pBTCT-C$_{12}$:PC$_{61}$BM bulk heterojunction solar cells. In contrast, extensive phase segregation in 1:4 blend devices leads to an efficient polaron generation resulting in an increased short circuit current density of the solar cell. 
For both studied ratios a bimolecular recombination of polarons was found using the complementary experiments. The charge carrier decay order of above two for temperatures below 300~K can be explained by a release of trapped charges. This mechanism leads to a delayed bimolecular recombination processes.
The experimental findings can be generalized to all polymer:fullerene blend systems allowing for phase segregation. 

\end{abstract}

This is the pre-peer reviewed version of the following article: Influence of phase segregation on the recombination dynamics in organic bulk heterojunction solar cells, A. Baumann, T. J. Savenije, D. H. K. Murthy, M. Heeney, V. Dyakonov, C. Deibel, which has been published in final form at \\ {\bf \href{http://onlinelibrary.wiley.com/doi/10.1002/adfm.201002358/abstract}{Adv.\ Func.\ Mat. 21, 1687 (2011)}}


\maketitle

\section{Introduction}
The power conversion efficiency (PCE) of organic solar cells recently reached 8~\%.\cite{green2011} 
The interpenetrating network of donor and acceptor phase is a key issue for further increasing the performance of polymer:fullerene bulk heterojunction solar cells, as it affects the exciton dissociation, charge transport and recombination.\cite{deibel2010review}
Whereas a fine phase intermixing on the sub-nm scale is believed to be beneficial for efficient photogeneration, the charge transport is strongly related to the percolation pathways formed within the bulk of the solar cell.
A crucial parameter having a great impact on the morphology of a bulk heterojunction solar cell is the donor--acceptor ratio. 
For many conjugated polymers, e.g. poly(p-phenylene vinylene) (PPV)~\cite{Shaheen2001}, a fullerene content of 67~wt.-\% to 80~wt.-\% was found to be optimal for the performance of those solar cells. Only a few polymer:fullerene systems have their optimum blend ratio at 1:1, e.g. poly(3-hexyl thiophene-2,5-diyl) (P3HT):[6,6]-phenyl-C$_{61}$-butyric acid methyl ester (PC$_{61}$BM).\cite{Chirvase2004}
Recently, it was reported that depending on the chain length and density of the polymer side chains, as well as the size of the fullerene, intercalation of the fullerenes in between the side chains can occur.\cite{Mayer2009,Cates2009} Mayer et al. found an optimum polymer:fullerene blend ratio of 1:3-1:4 in the  blend system consisting of poly(2,5-bis(3-tetradecyl thiophen-2-yl) thieno[3,2-b]thiophene) (pBTTT): phenyl-C$_{71}$-butyric acid methyl ester (PC$_{71}$BM), as an excess of PCBM molecules was necessary to create phase segregation. In MDMO-PPV:PCBM blend system the same optimum blend ratio of 1:3-1:4 was assigned to a nanoscale phase segregation with pure PCBM domains surrounded by a matrix of polymer containing up to 50~wt.-\% of fullerene.\cite{vanDuren2004}
Following an efficient charge generation, charge transport and recombination dynamics in blend systems are intuitively expected to be affected by the phase segregation and the dimensions of domain sizes, which will be addressed in detail in this work. 
Here we report on transport and recombination studies on poly(2,5-bis(3-dodecyl thiophen-2-yl) thieno[2,3-b]thiophene) (pBTCT-C$_{12}$):PC$_{61}$BM blend solar cells in which the ratio between the polymer and the fullerene was varied.
We used two complementary techniques, namely charge carrier extraction by linearly increasing voltage (photo-CELIV) probing the dynamics on a macroscopic level and time-resolved microwave conductivity (TRMC) for a microscopic point of view.
Note that both techniques yield information on the recombination processes very close to open circuit conditions, but on different length scales. 
Both techniques reveal an enhanced geminate recombination in the 1:1 weight ratio due to a fine donor--acceptor intermixing without phase segregation. In contrast, in the 1:4 ratio the formation of extended acceptor regions lead to efficient polaron pair dissociation.
The experimental findings are compared with the reference material system P3HT:PCBM.

\section{Results}

\begin{figure}
\centering
\includegraphics[width=0.6\linewidth]{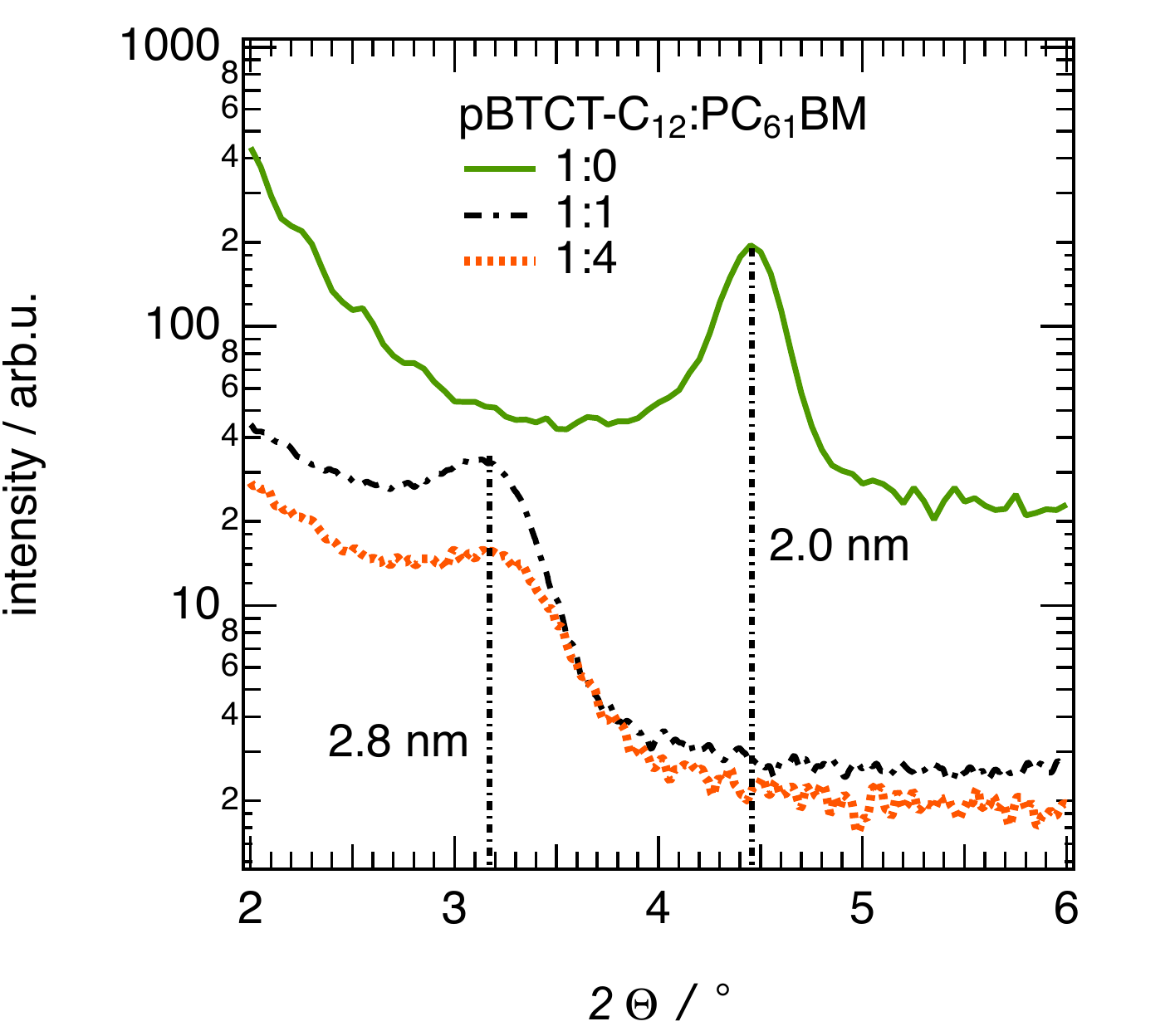}
\caption{\textbf{XRD on pure and blend films.} X-ray diffraction measurements on pure a pBTCT-C$_{12}$ film and blended with PC$_{61}$BM in a 1:1 and 1:4 ratio. The x-ray peak is shifting from 4.45$^{\circ}$ in the pristine pBTCT-C$_{12}$ film to 3.17$^{\circ}$ in the blend corresponding to a lattice distance of 2.0 nm and 2.8 nm, respectively.}
\label{fig:Fig1}
\end{figure}

Fig.~\ref{fig:Fig1} shows X-ray diffraction (XRD) measurements on as spun pBTCT-C$_{12}$ films and those blended with PC$_{61}$BM in a 1:1 and 1:4 weight ratio. No additional annealing was applied. For the pure pBTCT-C$_{12}$ film a diffraction peak at 4.3$^{\circ}$ is observed, corresponding to a lamellar spacing of 2.0~nm, i.e. the distance between two neighboring sheet-like structures consisting of $\pi$-stacked polymers.
Upon addition of PC$_{61}$BM, the X-ray peak was shifted towards smaller angles, corresponding to an increase in the lamellar packing distance to 2.8~nm. 
No further shift was observed from the 1:1 to the 1:4 weight ratio.

In order to study the macroscopic charge transport and recombination dynamics in the two blend ratios, we performed photo-CELIV measurements at temperatures ranging from $T=300$~K to $T=175$~K.

For both pBTCT-C$_{12}$:PC$_{61}$BM 1:1 and 1:4 devices charge carriers can be extracted from the bulk after laser excitation. However, considering the initial concentration of extracted charge carriers $n_{0}$ from the photo-CELIV measurements determined at a delay time of 150~ns, we observe an order of magnitude difference between the studied blend ratios. For pBTCT-C$_{12}$:PC$_{61}$BM 1:4 $n_{0}$ is 9.4$\times 10^{21}$~m$^{-3}$ at $T=300$~K, whereas for the 1:1 ratio the initial carrier concentration is decreased to a value of 9.6$\times 10^{20}$~m$^{-3}$.

\begin{figure}
\centering
\includegraphics[width=0.6\linewidth]{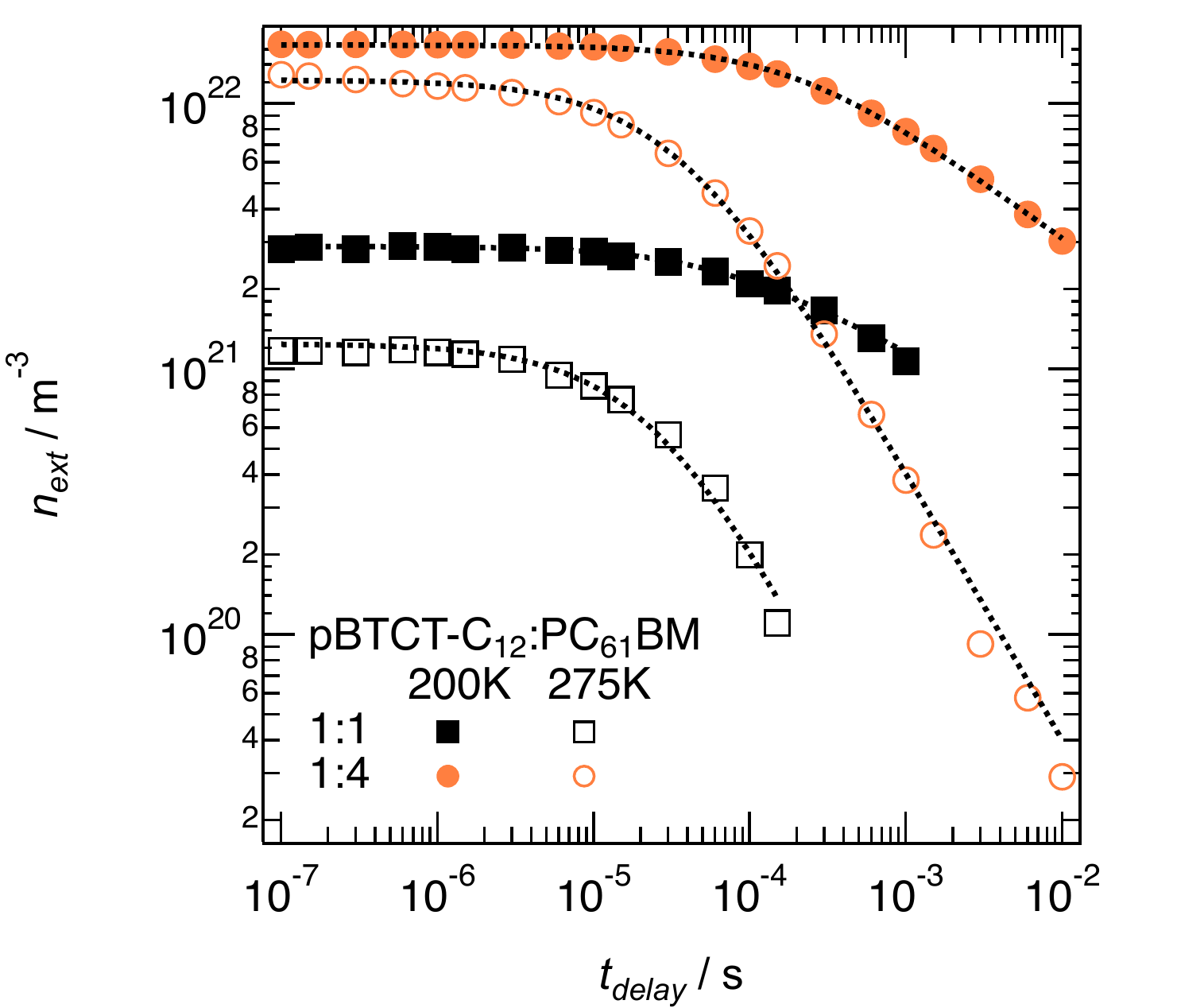}
\caption{\textbf{Photo-CELIV on 1:1 and 1:4 blend devices.} Extracted charge carrier density deduced from the photo-CELIV experiment at $T = 200$~K and $T=275$~K. The initial extracted charge carrier concentration (at $t_\text{delay}=100$~ns) is about an order of magnitude less for the 1:1 ratio compared to the 1:4 device. For $t_\text{delay}>100$~ns a bimolecular recombination with differing order can be observed for both ratios. pBTCT-C$_{12}$:PC$_{61}$BM 1:4 reveals a decay order of $\lambda+1=3.3$ at $T=200$~K and $\lambda+1=2.0$ at $T=275$~K, whereas in the 1:1 ratio an order of $\lambda+1=4.1$ and $\lambda+1=1.8$ can be determined for $T=200$~K and $T=275$~K.}
\label{fig:Fig2}
\end{figure}

In Fig.~\ref{fig:Fig2} the time dependent extracted charge carrier concentration $n_{ext}$ at $T=200$~K and $T=275$~K for pBTCT-C$_{12}$:PC$_{61}$BM 1:1 and 1:4 are compared.
For both ratios a bimolecular recombination process can be observed, but with different order of decay. At $T=275$~K the charge carrier decay is faster and more pronounced compared to $T=200$~K for 1:1 and 1:4 blend device, which can be seen in a stronger decrease of the carrier concentration at large delay times.

Using the generalized continuity equation $dn/dt = -kn^{\lambda+1}$---assuming that the spatial derivative of the current is zero at the built-in voltage---the experimental data can be fitted with a recombination coefficient $k$. The exponent corresponds to the order of the decay, $\lambda+1$.\cite{Shuttle2008a,Foertig2009} 
From the fits shown in Fig.~\ref{fig:Fig2} a charge carrier decay order of $\lambda+1=4.1$ for 1:1 pBTCT-C$_{12}$:PC$_{61}$BM at $T=200$~K can be obtained, whereas at $T=275$~K the order decreases to $\lambda+1=1.8$. For the 1:4 blend device, the recombination order decreases from 3.3 at $T=200$~K to $\lambda+1=2.0$ at $T=275$~K.

The average charge carrier mobility in the bulk deduced from photo-CELIV measurements at a fixed $t_{delay}$ of 60~$\mu$s is shown in Fig.~\ref{fig:Fig4}(a). The mobility in the 1:4 blend ratio is about an order of magnitude higher as compared to the 1:1 ratio over the entire studied temperature range with $\mu=4.8\times10^{-4}$~cm$^2$V$^{-1}$s$^{-1}$ for 1:4 and $\mu=4.5\times10^{-5}$~cm$^2$V$^{-1}$s$^{-1}$ for 1:1 pBTCT-C$_{12}$:PC$_{61}$BM BHJ solar cells at $T=300$~K. Note that no charge carriers could be extracted in 1:1 devices below $T=175$~K.

\begin{figure}
\centering
\includegraphics[width=0.6\linewidth]{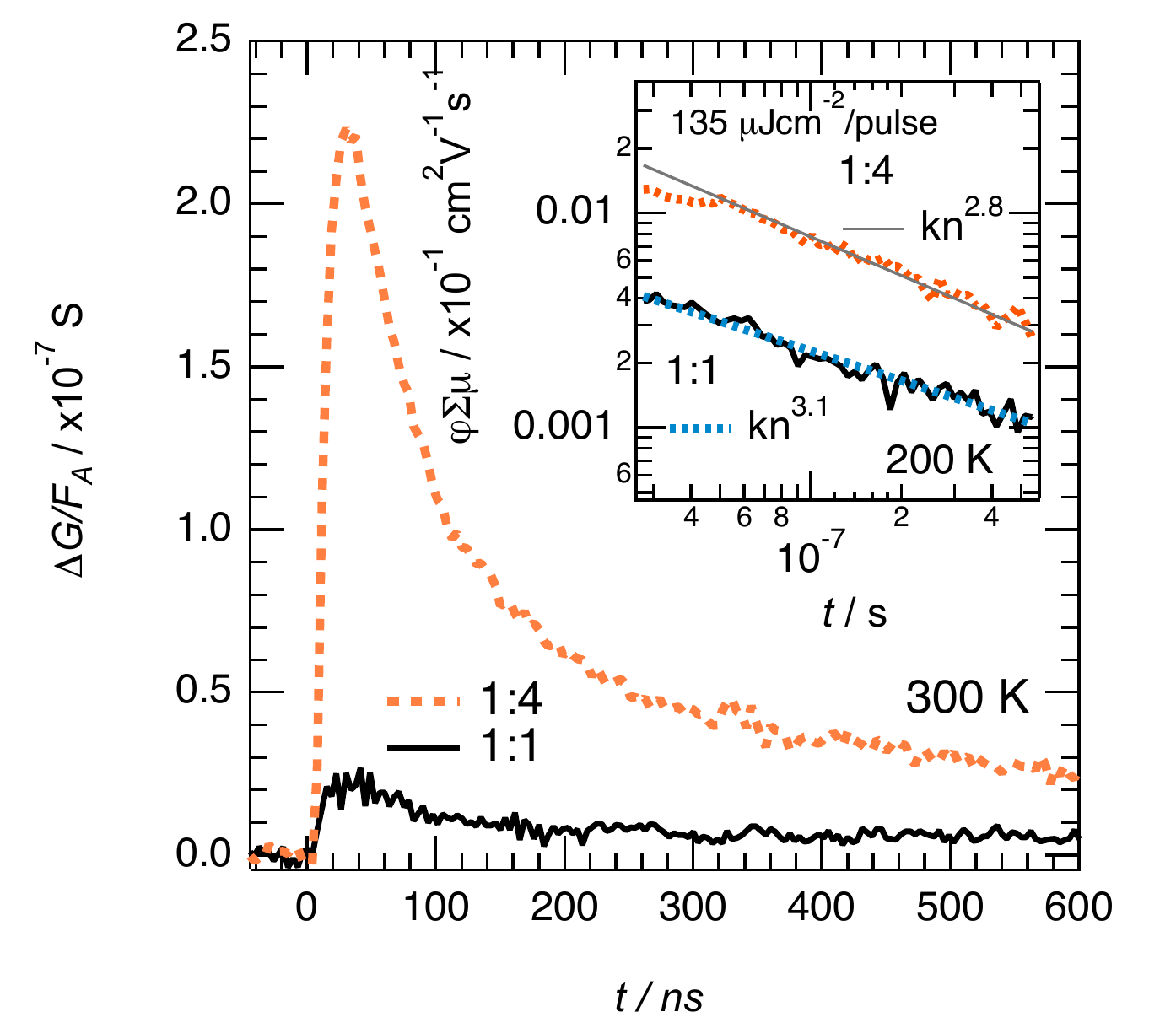}
\caption{\textbf{TRMC transients for 1:1 and 1:4 blends.} Photoconductance transients for the 1:1 (black) and 1:4 (orange) weight ratios of pBTCT-C$_{12}$:PC$_{61}$BM blends photoexcited at 532 nm and a fluence of 1.4$\times$10$^{14}$~photons cm$^{-2}$ per pulse at $T=300$~K. TRMC signals were normalized to the optical attenuation of the film. The inset shows the transients at $T=200$~K fitted by the generalized continuity equation. From the fits a decay order of 2.8 for the 1:4 blend device and 3.1 for the 1:1 ratio was determined. The laser intensity was adjusted to be 135~$\mu$Jcm$^{-2}$ per pulse.}
\label{fig:Fig3}
\end{figure}

\begin{figure}
\centering
\includegraphics[width=0.6\linewidth]{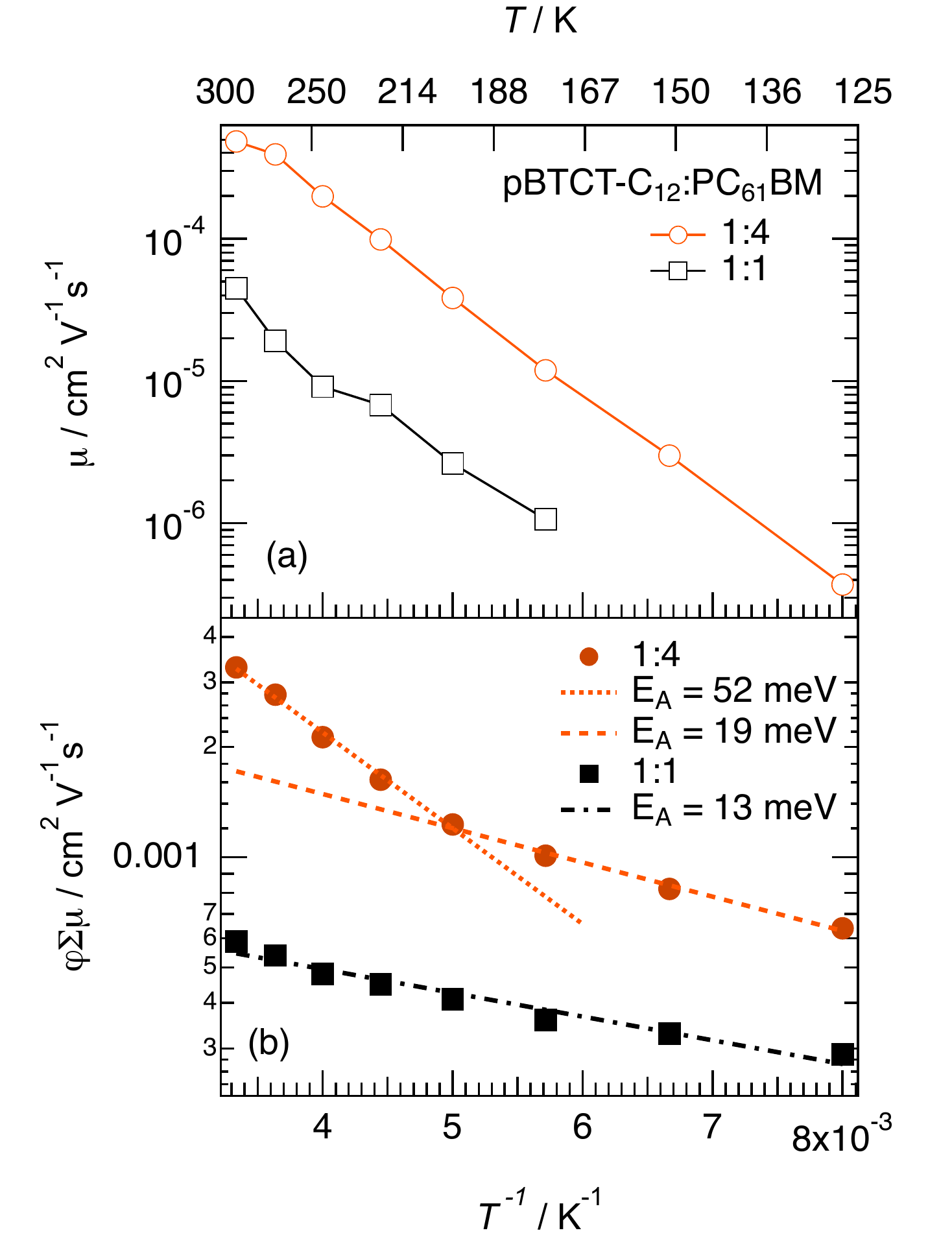}
\caption{\textbf{Temperature dependent charge carrier mobility.} (a) Average bulk charge carrier mobility extracted from the photo-CELIV transients at a fixed delay time of 60 $\mu$s and (b) $\varphi \Sigma \mu$ deduced from TRMC experiment is plotted against $1/T$ for pBTCT-C$_{12}$:PC$_{61}$BM 1:1 and 1:4. From TRMC two activation energies can be observed in the 1:4 blend ratio---52~meV for $T>200$~K and 19~meV for $T<200$~K---, whereas in the 1:1 ratio only one activation energy of 13~meV over the whole temperature range is detected}
\label{fig:Fig4}
\end{figure}

Furthermore, TRMC measurements were performed to investigate the local charge transport in pBTCT-C$_{12}$:PC$_{61}$BM 1:1 and 1:4 samples. TRMC signals of both samples show a fast initial rise due to the nanosecond laser pulse until the signal decays again. For the 1:4 blend the maximum photoconductivity $\Delta G_{MAX}$ at room temperature is more than one order of magnitude higher than that of the 1:1 blend (see Fig.~\ref{fig:Fig3}), which is in good agreement with the experimentally observed order of magnitude difference in the photo-CELIV signal for 1:1 and 1:4 devices. Variations of the photoconductivity on a similar scale depending on the relative PC$_{61}$BM concentration were observed previously for PC$_{61}$BM:poly(phenyl vinylene) derivatives.\cite{savenije2004}

The decrease of the photoconductivity after $\Delta G_{MAX}$ is due to charge recombination and  due to immobilization caused by trapping of charge carriers. Note that due to the absence of electrodes no charges are collected.
In the inset of Fig.~\ref{fig:Fig3} the product of the quantum efficiency for charge generation $\varphi$ and the sum of the mobilities of the positive and negative charge carrier $\Sigma\mu$ is shown for the 1:1 and 1:4 ratios for $T=200$~K.
For the TRMC transients we used the same fitting routine as for the photo-CELIV transients. In Fig.~\ref{fig:Fig4}(b) the product $\varphi \Sigma \mu$ is plotted as a function of $1/T$ for both studied blend ratios. Clearly, two different contributions are observed for the pBTCT-C$_{12}$:PC$_{61}$BM 1:4 sample. Below $T=200$~K a small activation energy of $E_{A}=19$~meV can be obtained, whereas  for temperatures above 200~K it becomes larger, i.e. $E_{A}=52$~meV. In contrast, only one activation energy of approximately $E_{A}=13$~meV can be seen in the 1:1 blend ratio.
At $T=300$~K the product $\varphi \Sigma \mu$ for the 1:1 weight ratio is $6\times 10^{-4}$~cm$^2$V$^{-1}$s$^{-1}$, whereas for the 1:4 blend ratio a value of $3.2\times 10^{-3}$~cm$^2$V$^{-1}$s$^{-1}$ can be determined.

\begin{figure}
\centering
\includegraphics[width=0.6\linewidth]{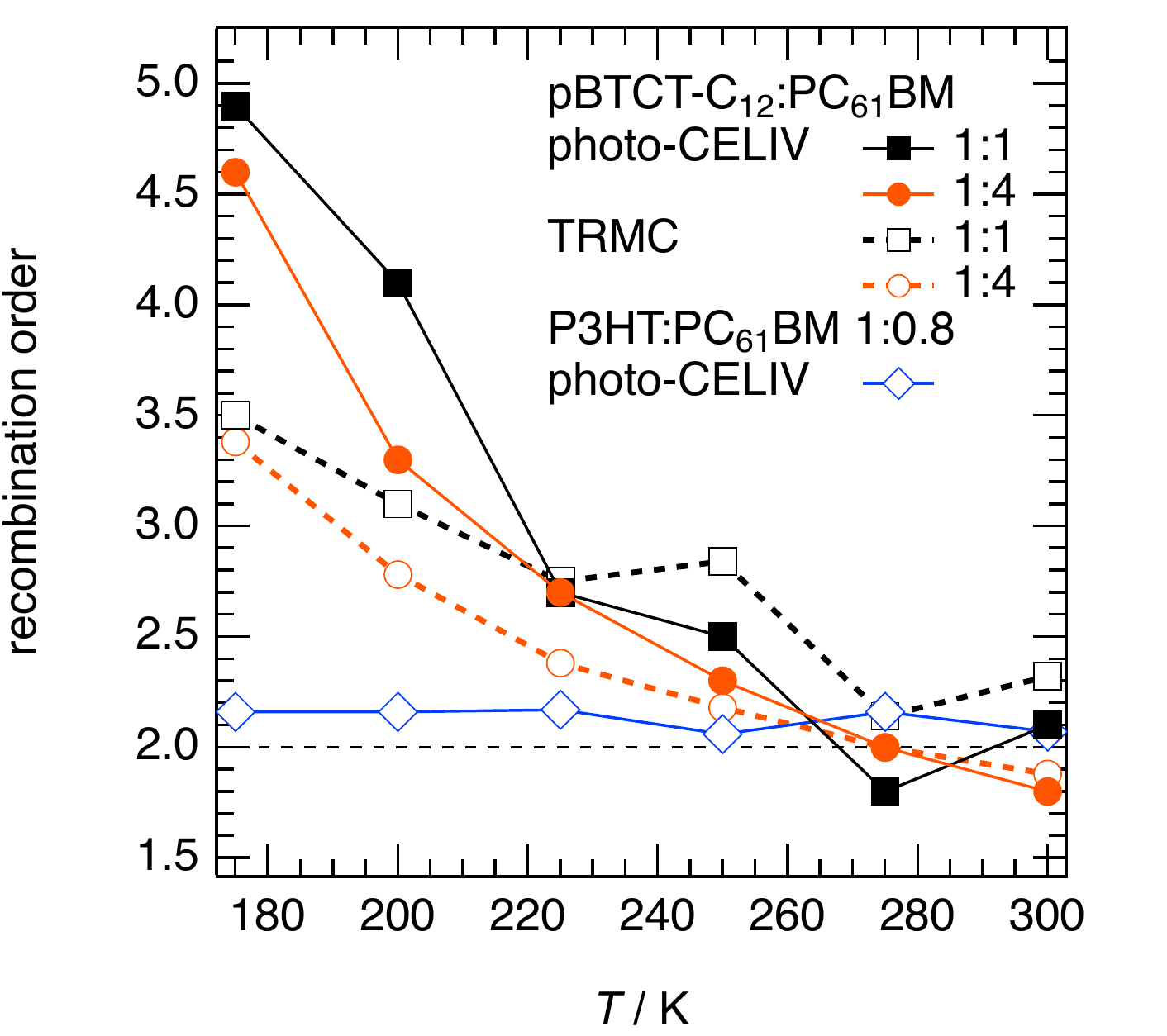}
\caption{\textbf{Charge carrier decay order observed in photo-CELIV and TRMC.} Recombination order is shown as a function of temperature deduced from fitting the $n_{ext}(t)$ data of the photo-CELIV and $\varphi\Sigma\mu$ data of the TRMC transients for pBTCT-C$_{12}$:PC$_{61}$BM 1:1 and 1:4. For comparison the blend system P3HT:PC$_{61}$BM is also shown.}
\label{fig:Fig5}
\end{figure}

The recombination order deduced from the fit of the photo-CELIV and TRMC transients for pBTCT-C$_{12}$:PC$_{61}$BM 1:1 and 1:4 BHJ solar cells is shown in Fig.~\ref{fig:Fig5} as a function of temperature.
From both techniques an increasing recombination order with decreasing temperature can be determined for both studied blend ratios. In the 1:4 device a charge carrier decay order of $\lambda+1=1.9$ was deduced from TRMC and $\lambda+1=1.8$ from photo-CELIV at $T=300$~K increasing to around 3.4 and 4.6 at $T=175$~K, respectively. Similar decay orders were observed in 1:1 blend devices with an increase from 2.1 to 4.9 obtained from the photo-CELIV experiments and 2.3 to 3.1 from corresponding TRMC transients. We note that slightly higher recombination orders are revealed from photo-CELIV measurements at low temperatures, the origin of which will be discussed. For comparison, the data of an annealed P3HT:PC$_{61}$BM 1:0.8 device is also shown. There, the experimentally determined recombination order increases only slightly from 2.1 to 2.2 in the studied temperature range from $T=300$~K to $T=175$~K, respectively.

\section{Discussion}


In X-ray diffraction measurements a lamellar spacing of 2.0~nm can be found in a pure pBTCT-C$_{12}$ film. This is slightly larger than the 1.84~nm spacing peak observed for a thermally annealed film,\cite{McCulloch2009} but is nonetheless shorter than the length of two dodecyl chains placed end-to-end, which is indicative of some interdigitation of the side chains of adjacent polymer backbones.\cite{McCulloch2009} For both pBTCT-C$_{12}$:PC$_{61}$BM blend ratios a clear shift of the XRD peak is observed pointing at an increase of the lamellar distance of 0.8~nm due to incorporation of methanofullerenes in the studied blend system.

Similar to PPV:PC$_{61}$BM blend systems\cite{vanDuren2004} it was suggested for pBTTT:PC$_{71}$BM devices---chemically related to the blend system studied in this work---that the addition of more than 50~wt.\% of fullerene lead to phase segregation with PC$_{71}$BM domains growing in proportion to an increasing amount of fullerene.\cite{Mayer2009}
Thus, although residual PC$_{61}$BM molecules in the 1:4 blends are not contributing to a further increase of the lamellar distance, the changed phase segregation does affect the transport properties and recombination dynamics in a bulk heterojunction solar cell, which was already suggested previously\cite{riedel2005,sanchez-diaz2010,shoaee2010} and will be discussed in the following.


In the 1:4 blend device a higher charge carrier mobility in the bulk (photo-CELIV) as well as detected locally (TRMC) is observed experimentally.
From temperature dependent TRMC measurements the contribution of charge carriers to the photoconductance can be deduced by determining the activation energy from an Arrhenius plot. In pBTCT-C$_{12}$:PC$_{61}$BM 1:4 two different activation energies could be assigned (see Fig.~\ref{fig:Fig4}) with a transition temperature of $T=200$~K.
Similar results were reported previously in different polymer:PC$_{61}$BM blend systems, e.g. P3HT:PC$_{61}$BM.\cite{Grzegorczyk2010} There, the activation energy for $T>200$~K was assigned to the contribution of electrons in PC$_{61}$BM, whereas at low temperatures the hole mobility in the polymer dominates the product $\varphi \Sigma \mu$.
If we project these findings to the here studied pBTCT-C$_{12}$:PC$_{61}$BM solar cells, above $T=200$~K we observe only an electron mobility in the 1:4 blends. In contrast, in the 1:1 blend ratio holes on pBTCT-C$_{12}$ dominate $\varphi \Sigma \mu$ over the entire studied temperature range. 
Note that from photo-CELIV measurements the polarity of the extracted charge carriers cannot be distinguished. 
However, we can conclude that in the 1:4 blend ratio extensive phase segregation leads to an efficient charge transport, whereas both charge carriers contribute to the transport.

In contrast, a reduced charge carrier mobility is observed in the 1:1 pBTCT-C$_{12}$:PC$_{61}$BM BHJ solar cell, which we assign to the fine donor--acceptor intermixing with less percolation paths for the polarons to the corresponding electrodes.
Furthermore, from current--voltage characteristics of the two blend ratios we find an almost three times higher short circuit current density under AM1.5G illumination in the 1:4 blend ratio compared to the 1:1 ratio. 


The initial charge carrier concentration experimentally found within (TRMC experiment, Fig.~\ref{fig:Fig3}) and extracted from (photo-CELIV, Fig.~\ref{fig:Fig2}) the device for both pBTCT-C$_{12}$:PC$_{61}$BM blend ratios was about an order of magnitude lower at 150~ns in the 1:1 device as compared to 1:4 with both techniques. This was despite the higher polymer content and thus larger absorption coefficient at 532~nm in the former. We note that the TRMC signal was corrected for the optical attenuation (see equation~\ref{eq:G2}). Photoluminescence in intercalated bulk systems is quenched very efficiently,\cite{parmer2008,Savenije2010} indicating that almost every photogenerated singlet exciton is separated at a donor--acceptor interface creating a polaron pair. However, fine phase intermixing and disorganized percolation paths in pBTCT-C$_{12}$:PC$_{61}$BM 1:1 blends may lead to a less efficient separation of photogenerated polaron pairs as compared to the 1:4 ratio. Thus, we assign the reduced charge carrier extraction in the 1:1 device at $t=150$~ns to an enhanced geminate recombination---occurring on shorter time scales---due to lower delocalization of the charge carriers caused by a lack of percolation paths. Electrons may be trapped on isolated acceptor molecules, which cannot hop away from the interface to be dissociated. Consequently, those charge carriers are lost for the photocurrent.
During the delay time the recombination of charge carriers will change the flat band conditions of the solar cell resulting in a sweep out of some of the charges, in addition to recombination. This effect is proportional to the charge carrier mobility and therefore stronger at higher temperatures. Therefore, a lower $n_{0}$ is observed in photo-CELIV measurements at high temperatures for a given donor--acceptor ratio (see Fig.~\ref{fig:Fig2}).

In the following, we will discuss the charge carrier decays after 150~ns in view of polaron recombination.
The crucial role of charge trapping for recombination orders above two will be discussed.

Recombination of charges in low mobility materials is usually described by the bimolecular Langevin process.\cite{langevin1903,pope1999book} This mechanism shows a charge carrier decay of the second order if electron and hole concentrations are similar. 

In typical donor--acceptor blends, the transport of electrons is restricted to the acceptor and the hole transport to the donor domain. In these systems, recombination of electrons and holes is spatially restricted to the heterointerface,\cite{koster2006} which explains in part the overestimation of recombination rates by applying the Langevin model.\cite{deibel2009}
In addition to reducing the overall recombination probability, it is important to note that the occurrence of non geminate recombination described by Langevin theory requires either mobile electrons and holes, $p_c$ and $n_c$ respectively, to meet at the heterointerface, or mobile electrons (holes) and holes (electrons) trapped very close to the heterointerface. 
Thereby, we distinguish between spatially $n_{t,s}$ and energetically $n_{t,e}$ trapped carriers.
The former is related to charge carriers being spatially confined in a material phase due to the lack of percolation to the respective electrode. In contrast to energetically trapped charges, they are locally mobile and can thus recombine at the heterointerface with an oppositely charged carrier, but cannot be extracted at the contacts. 

Energetically trapped charges---which are immobile---can only participate in the recombination process if (i) they are close enough to the donor--acceptor interface (for instance due to donor--acceptor intermixing), or (ii) if they are emitted after a certain dwell time, thus being mobile once again ($n_{t,e} \rightarrow n_{c}$). For (ii), the emission rate $e_t$ of charges from traps depends crucially on the energetic distribution of the density of trap states. In the hopping systems under consideration,
even the intrinsic density of states below the transport energy level---the tail of an exponential or a Gaussian density of states distribution (DOS)---can act as traps.\cite{arkhipov2001a} 

If we consider the continuity equation of free holes $p_c$, which can be extracted, we find $dp_c/dt=-kp_cn_c - k'p_cn_t$. Here, $n_c$ and $n_t$ are the free and trapped electrons, respectively, and $k$ and $k'$ are the relevant recombination prefactors. If $n_t \gg n_c, p_c$, the recombination dynamics become $dp_c/dt\approx- k'p_cn_t = p_c/\tau$, where $\tau=1/k'n_t\approx const$---a first order process. Depending on the magnitude of $n_t$ as compared to the mobile charge carriers, recombination orders between one and two would be observed.
In the studied pBTCT-C$_{12}$:PC$_{61}$BM blend system a recombination order far above two is observed at low temperatures by photo-CELIV and TRMC for both blend ratios (Fig.~\ref{fig:Fig5}) approaching close to or even decrease below two at $T=300$~K.
In 1:4 blends, the excess amount of fullerenes leads to the formation of acceptor rich domains, which allows a faster delocalisation on short time scales---improving the photogeneration yield~\cite{deibel2009a}---and to percolation paths for the electrons, favoring the charge extraction process as compared to the 1:1 blend. 
Due to fine intermixing in the 1:1 ratio, electrons will be mostly trapped, if not energetically, then spatially due to the lack of percolation paths on isolated fullerenes close to the polymer backbones. A large amount of polaron pairs recombines before dissociation, as indicated by the low initial signal magnitude for both, TRMC (Fig.~\ref{fig:Fig3}) and photo-CELIV (Fig.~\ref{fig:Fig2}) measurements. The remaining charge carriers might find percolation paths to be extracted.
However, with less thermal energy available they will partly be trapped in deep tail states of the DOS, thus being immobile. As emission is thermally activated, the release of trapped charges takes much longer at low temperatures. Only after emission from the continuous distribution of trap states, the now mobile charges can either be extracted by the linear voltage pulse or recombine non geminately. The latter process becomes less probable the later the emission occurs: the recombination partners of opposite charge may already have recombined or left the device. 
This leads to a reduced charge carrier decay. Therefore, we assign the observed high recombination order at low temperatures to a delayed release of trapped charges not actively participating in the recombination process.\cite{arkhipov1982,zaban2003,Foertig2009} 
With increasing temperature energetic trapping is more unlikely. However, the spatial restrictions in the bulk structure will affect the charge dynamics more dominantly.
The recombination order decreases even slightly below two in both studied ratios, which we attribute to an imbalanced relation between free mobile electrons and holes due to spatial trapping in the fine intermixed phase, as explained above. 

Even though the two complementary techniques photo-CELIV and TRMC rely on different measurement principles, the results fit qualitatively together. In case of TRMC, all mobile charge carriers contribute to the measurement signal, including also spatially restricted charge carriers $n_{t,s}$.
Photo-CELIV is only sensitive to charge carriers, which can be extracted by the voltage pulse. This includes also charge carriers being energetically trapped in deep states during the delay time not participating in the recombination processes, but can be extracted by the voltage pulse. Thus a higher recombination order is deduced from the photo-CELIV experiment.

\section{Conclusion}

The transport and recombination dynamics in pBTCT-C$_{12}$:PC$_{61}$BM bulk heterojunction solar cells were investigated by the combination of the two complementary techniques of photo-CELIV and TRMC for the first time probing different length scales. X-ray diffraction measurements indicate that structural changes occurred in the studied blend structure for a 1:1 and 1:4 weight ratio, where PC$_{61}$BM molecules probably intercalate into the available space of two neighboring polymer side chain stacks. 
Due to fine donor--acceptor intermixing, an enhanced geminate recombination is explained by the order in magnitude less concentration of initially extracted polarons in the 1:1 ratio. The lack of percolation pathways for electrons in the 1:1 ratio results in a low local charge carrier mobility decreasing the polaron pair dissociation probability significantly.
In contrast, in the 1:4 ratio extensive phase segregation lead to an efficient charge generation as the polarons can easily hop away from the heterointerface.
The polaron recombination dynamics studied by TRMC and photo-CELIV are bimolecular with an increasing order with decreasing temperature for both studied ratios, which is due to charge trapping.

\section{Experimental}

BHJ solar cells were prepared by spin coating a 35 nm layer of poly(3,4-ethylene dioxy thiophene):poly(styrene sulfonate) (Baytron~P~VP~AI~4083) on indium tin oxide samples with post-annealing step of 130$^{\circ}$C for 10 minutes. The pBTCT-C$_{12}$:PC$_{61}$BM blends made from solutions of 16~mg/ml and 10~mg/ml in 1,2-ortho-dichlorobenzene were spin coated in an inert atmosphere with active layers of a thickness in the range of 70 to 280~nm.
Without further annealing the active blend layer, Ca (3 nm)/Al (100 nm) were evaporated thermally on top. PC$_{61}$BM was purchased from Solenne, poly(2,5-bis(3-dodecylthiophen-2-yl)thieno[2,3-b]thiophene) (pBTCT-C$_{12}$) was synthesized according to the published procedure in Ref.~\cite{Heeney2005}. The molecule weight was Mn 22,600~g/mol Mw 39,300~g/mol as determined by GPC in chlorobenzene at 60$^{\circ}$C on an Agilent 1100 series HPLC using two Polymer Laboratories mixed B columns in series, with calibration against narrow weight PL polystyrene calibration standards.
All materials were used without further purification. 

The current--voltage characteristics of the organic solar cells were measured in a nitrogen glovebox. We used an Oriel 1160 AM1.5G solar simulator for illumination.
For investigations of the charge transport and recombination studies using the photo-CELIV technique, the solar cells were transferred to a closed cycle helium cryostat with helium as inert contact gas.
A triangular voltage pulse in reverse direction is applied to the solar cells extracting free charge carriers from the bulk. When exciting the sample with a laser flash, photogenerated charge carriers can be studied.~\cite{mozer2005b} By applying an offset voltage to compensate for the built-in field of the solar cells and varying the delay time between the laser excitation and the charge carrier extraction, the recombination dynamics can be studied by measuring the time dependent charge carrier concentration. Using a parametric equation for the mobility~\cite{deibel2009c}
\begin{equation} 
\mu = \dfrac{2d^2}{3A't_{max}^2(1+0.21\dfrac{\Delta j}{j_0})}\quad{,}
\label{eq:eq1}
\end{equation}
the average charge carrier mobility in the bulk can be calculated from the transient, with $A'=V_{pulse}/t_{pulse}$, $d$ the sample thickness, $t_{max}$ the time of the transient maximum, $\Delta$j as the maximum of the extraction current and $j_{0}$ the current step related to the capacitive displacement current.\cite{Juska2000,Lorrmann2010}
A double pulse generator (Agilent 81150A) was used for applying a triangular voltage pulse to the solar cell. The current transients were acquired by a digital oscilloscope (Agilent Infiniium DSO90254A) after amplification by a current--voltage amplifier (FEMTO DHPCA-100).
Photo-CELIV measurements were performed at different temperatures ranging from 150~K to 300~K in steps of 25~K. The delay time between the laser excitation and the extraction voltage pulse was varied from 100~ns to 10~ms. 
We used the second harmonic of a Nd:YAG laser ($\lambda=532$~nm, $<$ 80~ps pulse duration) for laser excitation.

Furthermore, the decay of charge carriers was studied by transient microwave conductivity. By using this technique both mobile holes and electrons contribute to the measured change in microwave conductance ($\Delta$G). 

The microscopic charge carrier mobilities were deduced from the TRMC transients by equation
\begin{equation}
\varphi\Sigma\mu = \dfrac{\Delta G_{MAX}}{\beta e I_0 F_A}\quad{.}
\label{eq:G2}
\end{equation}
$\varphi$ is the quantum efficiency for charge generation and $\Sigma\mu$ is the sum of the mobilities of the positive and negative charge carrier. $\beta$ denotes the ratio between the broad and narrow inner dimension of the waveguide, $e$ is the elementary charge, $J_{0}$ the incident laser fluence and $F_{A}$ is the fractions of photons absorbed by the sample. The laser fluence in the TRMC experiment was 135~$\mu$Jcm$^{-2}$ per pulse.

As described in more detail previously~\cite{Kroeze2002,Kroeze2003,Huijser2005}, if $\varphi$ is undetermined, the value of $\varphi\Sigma\mu$ is a lower limit of the local charge carrier mobility.
Samples for TRMC measurements were prepared from the same solution as the solar cells.

\begin{acknowledgments} 

The current work is supported by the Bundesministerium f{\"u}r Bildung und Forschung in the framework of the GREKOS project (contract no.~03SF0356B). A.B. thanks the German Federal Environmental Foundation (Deutsche Bundesstiftung Umwelt, DBU) for funding. The work of D.H.K.M. forms part of the research programme of the Dutch Polymer Institute (DPI), project Nr. 681. C.D. gratefully acknowledges the support of the Bavarian Academy of Sciences and Humanities. V.D.'s work at the ZAE Bayern is financed by the Bavarian Ministry of Economic Affairs, Infrastructure, Transport and Technology. 

\end{acknowledgments}

\end{document}